\documentclass{jaa}
\usepackage{graphicx,graphics}
\usepackage{amssymb}
\usepackage{amsmath}
\usepackage{epsfig}
\usepackage{float}
\usepackage{array}
\usepackage{comment}
\usepackage{color}
\usepackage[graphicx]{realboxes}

\begin{document}\sloppy

\title{Prospects of detecting Fast Radio Bursts using Indian Radio Telescopes}

\author{Siddhartha Bhattacharyya\textsuperscript{1,*}}
\affilOne{\textsuperscript{1}Department of
Physics, Indian Institute of Technology, Kharagpur, India\\}

\twocolumn[{

\maketitle

\corres{siddhartha@phy.iitkgp.ac.in}

\msinfo{DD MMMM YYYY}{DD MMMM YYYY}

\begin{abstract}
Fast Radio Bursts (FRBs) are short duration highly energetic dispersed radio pulses. We developed a generic formalism (Bera {\em et al.}, 2016) to estimate the FRB detection rate for any radio telescope with given parameters. By using this model, we estimated the FRB detection rate for two Indian radio telescope; the Ooty Wide Field Array (OWFA) (Bhattacharyya {\em et al.}, 2017) and the upgraded Giant Metrewave Radio Telescope (uGMRT) (Bhattacharyya {\em et al.}, 2018) with three beam forming modes. In this review article, I summarize these two works. We considered the energy spectrum of FRBs as a power law and the energy distribution of FRBs as a Dirac delta function and a Schechter luminosity function. We also considered two scattering models proposed by Bhat {\em et al.} (2004) and Macquart \& Koay (2013) for these work and I consider FRB pulse without scattering as a special case for this review. We found that the future prospects of detecting FRBs by using these two Indian radio telescopes is good. They are capable to detect a significant number of FRBs per day. According to our prediction, we can detect $\sim 10^5-10^8$, $\sim 10^3-10^6$ and $\sim 10^5-10^7$ FRBs per day by using OWFA, commensal systems of GMRT and uGMRT respectively. Even a non detection of the predicted events will be very useful in constraining FRB properties.
\end{abstract}


}]

\doinum{12.3456/s78910-011-012-3}
\artcitid{\#\#\#\#}
\volnum{000}
\year{0000}
\pgrange{1--}
\setcounter{page}{1}
\lp{1}

\section{Introduction}
Fast Radio Bursts (FRBs) are short duration ($\sim$ ms), highly energetic ($\sim 10^{32} - 10^{34}$ J) dispersed radio pulses, first discovered (Lorimer {\em et al.}, 2007) at the Parkes radio telescope. The high dispersion measure (DM) of the detected FRBs, which is in general $\sim 5-20$ times excess DMs compare to what is expected from the Milky Way (Cordes \& Lazio, 2003), strongly suggests that FRBs are extragalactic events. The observed dispersion and the scattering indices imply the fact that the FRB signal propagates through the cold ionized plasma (Katz, 2016) of the interstellar medium (ISM) of the Mliky Way, host galaxy of the source and the intergalactic medium (IGM). A total $35$ FRBs have been reported\footnote{http://frbcat.org/} to date, of these $26$ FRBs have been detected at the Parkes radio telescope(Petroff {\em et al.} (2016); Petroff {\em et al.} (2017); Keane {\em et al.} (2016); Ravi {\em et al.} (2016); Bhandari {\em et al.} (2017); Shannon {\em et al.} (2017); Price {\em et al.} (2018), Oslowski {\em et al.} (2018a) \& Oslowski {\em et al.} (2018b)), six FRBs have been detected at the UTMOST radio telescope (Caleb {\em et al.} (2017); Farah {\em et al.} (2017) \& Farah {\em et al.} (2018)) and one each has been detected at the Arecibo (Spitler {\em et al.}, 2014), GBT (Masui {\em et al.}, 2015) and ASKAP (Bannister {\em et al.}, 2017) radio telescopes. One FRBs has been found to repeat (Scholz {\em et al.}, 2016) and $17$ detections from the same source have been reported to date. There are several models (Kulkarni {\em et al.}, 2015) proposed for the emission mechanism of FRBs but the exact one is still unknown. The energy spectrum and the energy distribution of FRBs are not well constrained and moreover the estimates of the spectral index of FRBs are available only for few FRBs but they are not reliable estimated due to the poor localization of the source within the single dish beam. 

We (Bera {\em et al.}, 2016) developed a generic formalism to estimate the detection rate and the redshift distribution of FRBs for a radio telescope with given parameters. We assumed a power law $E_{\nu}\propto\nu^{\alpha}$ with $\alpha$ as the spectral index for the energy spectrum of FRBs and two scattering models proposed by Bhat {\em et al.} (2004) and Macquart \& Koay (2013) for the predicted pulse width of FRBs and they are denoted here as scattering model I (Sc-I) and II (Sc-II) respectively. Scattering model I (Sc-I) is an empirical fit to a large number of pulsar data in the Milky Way, whereas scattering model II (Sc-II) is purely theoretical without any observational consequences. The details mathematical expression of scattering model I \& II can be found in Bera {\em et. al.} (2016). In this review, I also consider FRB pulse without scattering as a special case, since for the most of the detected FRBs we did not find any scattering. The model is normalized by considering FRB $110220$ as the reference event and the estimated energy ($E_0=5.4\times10^{33}\,{\rm J}$) of this FRB using our model as the reference energy. 
In this review article, I consider the prescribed FRB rate from Champion {\em et al.} (2016), {\it i.e.} $7\times10^3$ FRBs per sky per day as the reference event rate. Note that this prescribed FRB rate is differed from the FRB rate that we used in our previous publications (Bear {\em et al.}, 2016; Bhattacharyya {\em et al.}, 2017 \& Bhattacharyya {\em et al.}, 2018) by factor of $\sim 5\times10^5$.
Note that, the value of $E_0$ is estimated by using the model prescribed in Bera et al. (2016) with $\alpha=-1.4$ ($E_{\nu}\propto\nu^{\alpha}$), which is differed from the energy mentioned in Thornton et al. (2013) by a factor of $5$.
As described in (Bera {\em et al.}, 2016), all redshifts are inferred from the DM; the scattering time scale, when available gives an upper limit on the redshift.  We also considered two energy distribution functions, a Dirac delta function and a Schechter luminosity functions with the exponent in the range $-2\leq\gamma\leq 2$, as the possible energy distribution functions of FRBs.

Using the model mentioned above, we estimated the FRB detection rate for the two Indian radio telescopes, Ooty Wide Field Array (OWFA) (Bhattacharyya {\em et al.}, 2017) and the upgraded Giant Metrewave Radio Telescope (uGMRT) (Bhattacharyya {\em et al.}, 2018). We have found that the detection probability of FRBs largely depends on two factors, the field-of-view (FoV) of the telescope and the antenna sensitivity (${\rm A}_{\rm S}$), where the antenna sensitivity is the ratio of the antenna gain (G) and the system temperature (${\rm T}_{\rm sys}$) of the telescope. A telescope with large field-of-view (FoV) and high antenna sensitivity (${\rm A}_{\rm S}$) is capable of detecting a large number of FRBs. Hence, the product FoV $\times$ ${\rm A}_{\rm  S}$ is a measure of the FRB detection sensitivity for a telescope. The typical value of this product for OWFA with the observational frequency of $326.5\;{\rm MHz}$ and uGMRT with the observational frequency of $375\;{\rm MHz}$ are $1.054$ and $1.63\times10^{-2}\;{\rm deg}^2\, {\rm  Jy}^{-1}$ respectively. However, this product is estimated by considering the higher Galactic latitude ({\em i.e.} cold sky). For comparison, this value is $1.96\times10^{-2}\;{\rm  deg}^2\,{\rm Jy}^{-1}$ for the Parkes telescope and $3.93\times10^{-4}\;{\rm deg}^2\,{\rm Jy}^{-1}$ for the Arecibo telescope. In this respect both OWFA and uGMRT are capable to detect a large number of FRBs in compare to the Parkes and Arecibo radio telescopes.

This paper is a review of our previous works. I summarized our predictions of FRB detection rates for OWFA and uGMRT. A brief outline of the paper is as follows. Section 2 presents a brief description and the FRB detection rates for OWFA. Scetion 3 presents a brief description and the FRB detection rates for GMRT and uGMRT. Section 4 presents 
FRB detection rate and localization comparison among OWFA, commensal systems of GMRT and uGMRT. Finally, I discuss and summarize the results in Section 5.

\section{Ooty Wide Field Array (OWFA)}
The Ooty Wide Field Array (OWFA) is an upgraded version of the Ooty Radio Telescope (ORT) which was built in early 70's (Swarup et al. 1971) at Ooty, Tamil Nadu. ORT has a long cylindrical reflector of dimension $530{\rm m}\times 30{\rm m}$,which contains $1056$ half wavelength linear dipoles along the focal line of the reflector. 
The signal from the dipoles can be combined in different way. Currently the signals from these dipoles are combined to form an analogue incoherent beam forming network which we referred as the Legacy System. The Legacy System (LS) operates at an observational frequency $\nu_0=326.5\,{\rm MHz}$ ($\lambda=0.91\,{\rm m}$) with bandwidth $B=4\,{\rm MHz}$.
The system is being upgraded to two modes of operation; Phase I (PI) and Phase II (PII). In Phase I, $24$ dipoles combined together to form a single element and this system has a total $40$ such elements. In Phase II, $4$ dipoles combined together to form a single element and this system has a total $264$ such elements. The bandwidth of Phase I and Phase II are $19.2\,{\rm MHz}$ and $38.4\,{\rm MHz}$ respectively, centred at the same observational frequency of the ORT Legacy System. More technical information about OWFA can be found in Subrahmanyan et al. (2016). The system specifications of LS, PI and PII are tabulated in Table 1.

In this work, we considered three kind of beam formations; incoherent (IA), coherent single (CA-SB) and coherent multiple (CA-MB) beam formations. In the case of incoherent beam formation (IA), the squares of the voltages from the individual elements are summed over to obtain the total power. This mode of beam formation does not contain any phase information. Here the FoV is proportional to $\lambda/d$, where $d$ is the length of a single element. The sensitivity in this mode is increased by a factor of $\sqrt{N_A}$ compared to the sensitivity achieved by a single element. Note that LS operates with incoherent beam forming mode.
In the case of coherent single beam formation (CA-SB), the voltage signals from the individual elements with phase are added directly and then squared to obtain total power. In this mode, the field of view (FoV) is proportional to $\lambda/D$, where $\lambda$ is the wavelength of the observation and $D$ is the length of the largest baseline. Here the sensitivity is increased by a factor of $N_A$ compared to the sensitivity achieved by a single element. The coherent multiple beam formation is a mixture of IA and CA-SB mentioned above. 
In coherent multiple beam formation (CA-MB), one forms the IA to obtain a large instantaneous field of view but at a relatively shallow sensitivity. When an event is detected in the IA mode, the high time resolution signals are recorded to eventually form multiple coherent beams offline in all possible directions. This will give us the sensitivity of the CA-SB, but with the field of view of the IA and hence the detection probability in this mode is larger than the same for IA and CA-SB modes. This specific kind of strategy was first demonstrated in a pilot transient survey with the Giant Metrewave Radio Telescope (GMRT) by Bhat et al. (2013).
Note that the value of threshold signal to noise ratio ($({\rm S}/{\rm  N})_{\rm th}$) for this beam forming mode is less in compare to the same for IA and CA-SB modes. In this work, $({\rm S}/{\rm  N})_{\rm th}=3$ for CA-MB mode and $({\rm S}/{\rm  N})_{\rm th}=10$ for IA and CA-SB beam forming modes.
\begin{figure}
\centering
\includegraphics[width=\columnwidth]{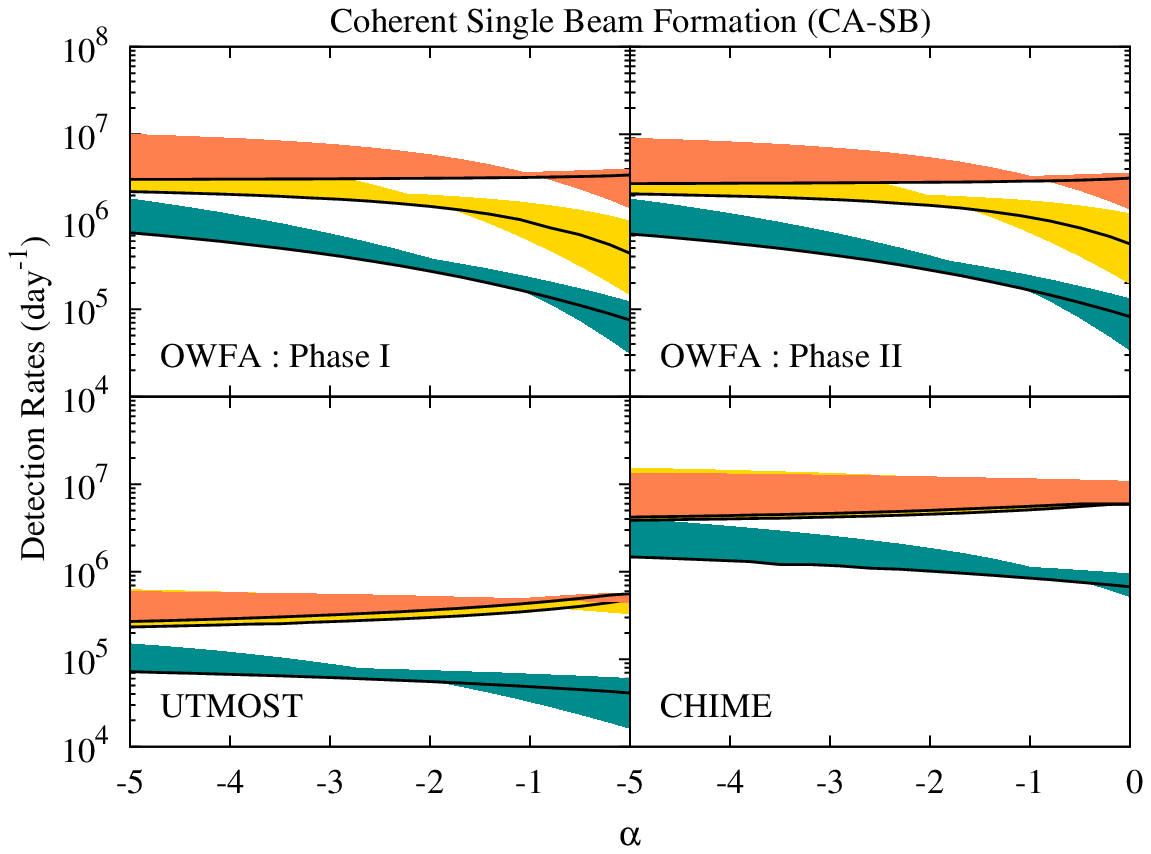}
\includegraphics[width=\columnwidth]{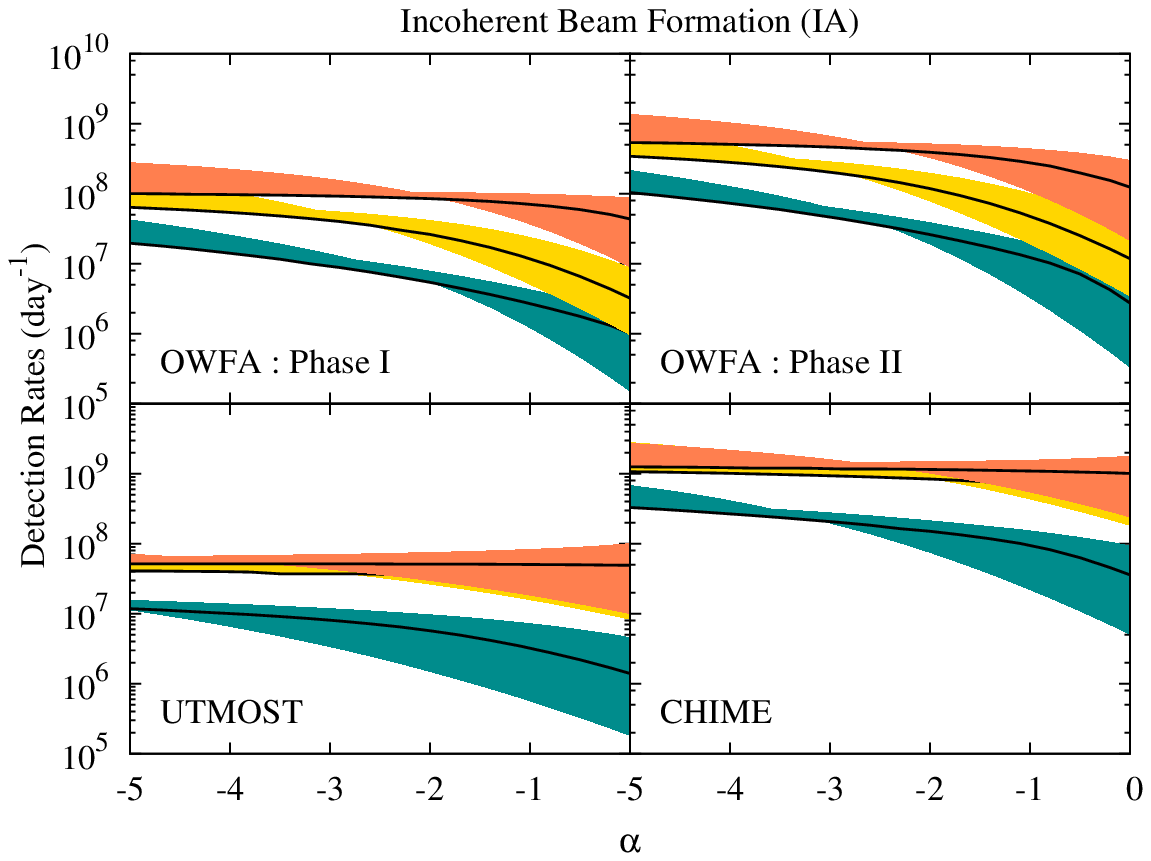}
\includegraphics[width=\columnwidth]{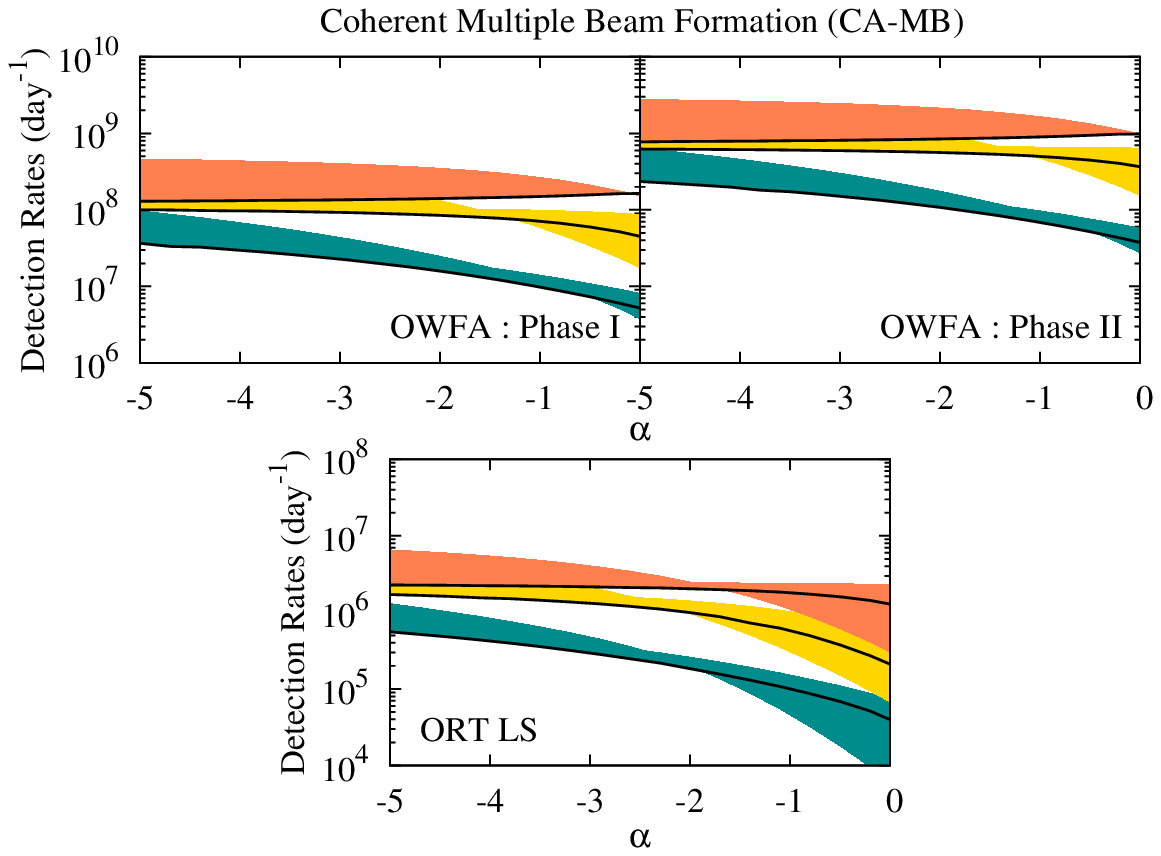}
\caption{The variation of the FRB detection rate with the variation of $\alpha$, where $E_{\nu}\propto\nu^{\alpha}$, for ORT legacy system, OWFA Phase I and Phase II, UTMOST and CHIME with three kind of beam formations (IA, CA-SB \& CA-MB). 
Two scattering models Sc-I (green region) and Sc-II (yellow region) and FRB pulse without scattering (orange region) have been considered here. The solid  black lines denote the Dirac delta function, while the boundaries of the regions enclose the curves correspond to the Schechter luminosity function with an exponent in the range $-2\leq\gamma\leq2$.}
\label{fig:1}
\end{figure}

We considered the energy spectrum of FRBs as $E_{\nu}\propto\nu^{\alpha}$ for this work and $\alpha$ is defined here as the spectral index. We estimated the FRBs detection rates for ORT-legacy system, OWFA Phase I and II and compared the results with other two cylindrical radio telescopes, UTMOST (Caleb {\em et al.} 2016) and CHIME (Newburgh {\em et al.} 2014). Note that UTMOST operates at an observational frequency of $843\;{\rm MHz}$ with a bandwidth of $31.25\;{\rm MHz}$, whereas CHIME operates at an observational frequency of $600\;{\rm MHz}$ with a bandwidth of $400\;{\rm MHz}$.
Figure \ref{fig:1} shows the FRB detection rates as a function of $\alpha$ for OWFA Phase I \& II, UTMOST and CHIME with three beam forming modes IA, CA-SB and CA-MB and the ORT legacy system. It is found that the detection rate varies with different scattering models and the detection rate is maximum for the case of FRB pulse without scattering, which is roughly two order and one order of magnitude larger than the same for scattering model I and II respectively. Further the detection rate increases with decreasing $\alpha$ ($\alpha\leq 0$). In a brief we expect to detect $\sim 10^5-10^8$ FRBs per day by using OWFA Phase II with fluence $F\geq0.7\;{\rm Jy\,ms}$, which is also large in compare to the that for UTMOST. However, the detection rate is maximum for the CHIME due to its large field of view but these three telescopes operate in different frequency ranges and hence complementary. 

\section{upgraded Giant Metrewave Radio Telescope (uGMRT)}
The GMRT antennas are distributed in a Y shaped pattern with a shortest baseline of $200$ m and a longest baseline of $25$ km. Each dish has five prime focus feeds, only one of which is available at a given time, having five discrete operational frequencies centered at $150\,{\rm MHz}$, $235\,{\rm MHz}$, $325\,{\rm MHz}$,  $610\,{\rm MHz}$ and $1280\,{\rm MHz}$ with a maximum backend instantaneous frequency bandwidth of $32\,{\rm MHz}$.  Currently the GMRT is going through an upgradation (Gupta {\em et al.}, 2017), to provide significantly large instantaneous bandwidths with four operational frequencies, \textit{viz.} Band $2$ at $185\,{\rm MHz}$ with a bandwidth of $130\,{\rm MHz}$, Band $3$ at $375\,{\rm MHz}$ with a bandwidth of $250$ MHz, Band $4$ at $700\,{\rm MHz}$ with a bandwidth of $300\,{\rm MHz}$ and Band $5$ at $1250\,{\rm MHz}$ with a bandwidth of $400\,{\rm MHz}$.

In this work we also considered the proposed commensal system for the GMRT (Bhattacharyya {\em et al.} 2018). This system as currently envisaged, would reuse the legacy signal transport chain of the GMRT, which has a bandwidth of $32\,{\rm MHz}$.  The feed system is proposed to be mounted off-focus on the quadripod feed legs of the GMRT, and hence be available at all times, unlike the main feeds, which are mounted on a rotating turret, and of which only one feed is available at a given time. We examined the expected detection rate for two possible central frequencies, viz. $300\,{\rm MHz}$ and $450\,{\rm MHz}$ with a bandwidth of $32\,{\rm MHz}$, which roughly span the possible frequencies of the proposed system and they are denoted here as  Bands $S1$ and $S2$ respectively. In the both GMRT commensal systems and uGMRT, we considered three kind of beam formations; incoherent (IA), coherent single (CA-SB) and multiple incoherent (MIA) beam formations. The description of IA and CA-SB have been mentioned earlier and MIA mode is a special beam forming mode for the both commensal systems of GMRT and the four frequency bands of uGMRT.

In the MIA beam forming mode, the entire array is divided into multiple ($N_{\rm Array}$) sub-arrays  each of which operates in the IA mode. This will give us the large field-of-view of the IA with a shallow sensitivity compared to IA.  A signal is considered as an event if and only if it is detected in all the sub arrays.  In practice, because the co-incidence filtering greatly reduces false alarms (Bhat {\em et al.}, 2013), one can use a lower signal to noise ratio threshold $({\rm S}/{\rm  N})_{\rm th}$ ($\geq3$) for each sub-array. Although the MIA mode has a lower sensitivity compared to the IA mode, the reduced detection threshold more than compensates for this and we found a higher FRB detection rate for the MIA mode as compared to the IA mode. Note that in this work, we considered $({\rm S}/{\rm  N})_{\rm th}=3$ and $N_{\rm Array}=3$ for the MIA beam forming mode. The system specifications of two commensal systems of GMRT and four frequency bands of uGMRT are tabulated in Table 1.

\begin{figure}
\centering
\includegraphics[width=\columnwidth]{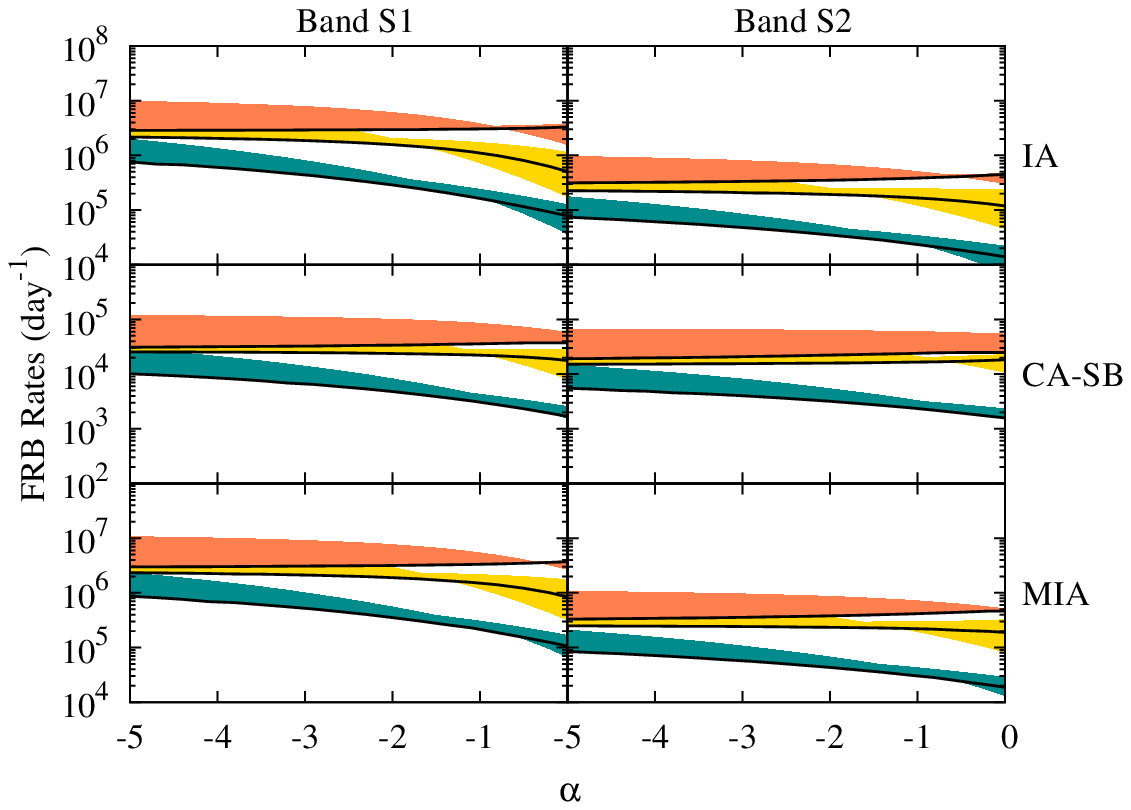}
\includegraphics[width=\columnwidth]{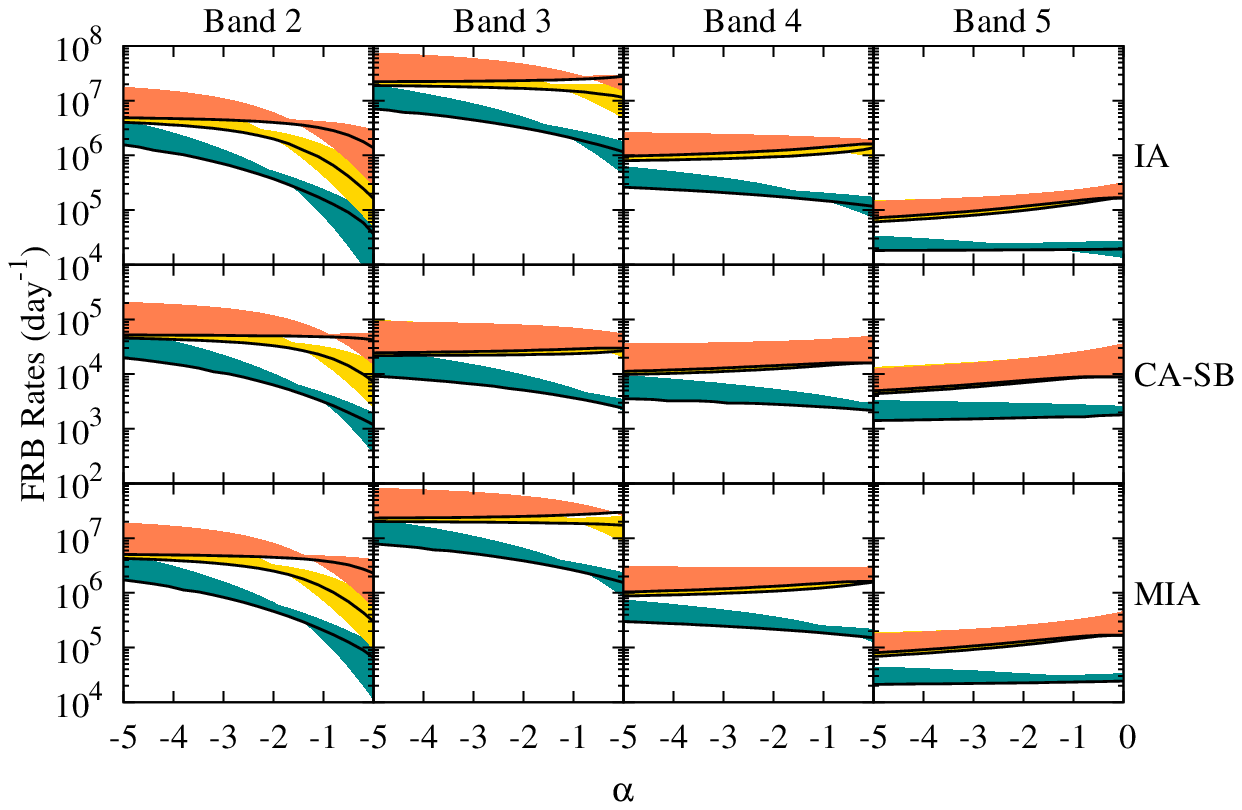}
\caption{The variation of the FRB detection rate with the variation of $\alpha$ for the proposed commesal system of GMRT and the four frequency bands of uGMRT with three kind of beam formations (IA, CA-SB \& MIA). Two scattering models Sc-I (green region) and Sc-II (yellow region) and the FRB pulse without scattering (orange region) have been considered here.} The solid  black lines denote the Dirac delta function, while the boundaries of the regions enclose the curves correspond to the Schechter luminosity function with an exponent in the range $-2\leq\gamma\leq2$.
\label{fig:2}
\end{figure}

Figure \ref{fig:2} shows the variation of FRB detection rates with the variation of $\alpha$  for the commensal systems of GMRT and the four frequency bands of uGMRT with three beam forming modes (IA, CA-SB \& MIA). As I have mention earlier, it is also found here that the detection rate varies with different scattering models and the detection rate is maximum for the case of FRB pulse without scattering, which is roughly $\sim10$ and $\sim2$ times larger than the same for scattering model I and II respectively. Further it also increases with decreasing $\alpha$. In a brief, we expect to detect $\sim 10^3-10^6$ FRBs per day with fluence $F\geq0.41\;{\rm Jy\,ms}$ for the commensal system of GMRT and $\sim 10^5-10^7$ FRBs per day with fluence $F\geq0.12\;{\rm Jy\,ms}$ for the four frequency bands of uGMRT respectively.

\section{FRB detection rate and localization comparison}
In this section, I compare FRB detection rate and localization of the event for the three phases of OWFA (LS, PI \& PII), the two commensal systems of GMRT (Band $S1$ \& $S2$), the four observational frequency bands of uGMRT (Band $2$, $3$, $4$ \& $5$) and CHIME. Table 1 shows the number of FRBs expected to be detected per day by considering the Dirac delta function as the energy distribution function of FRBs and the mean value of $\alpha$ with error for the range $-5\leq\alpha\leq0$, where $E_\nu\propto\nu^{\alpha}$, for different systems of OWFA, GMRT, uGMRT and CHIME with their observational frequency, bandwidth, beam formations, field of view and threshold fluence of the event required for the detection. It is found that the detection rate is larger for FRB pulse without scattering than the same for scattering model I and II respectively. In Table 1, the detection rate is large ($\sim 8.26\times10^8$ FRBs per day) for OWFA Phase II with CA-MB beam forming mode in comparison to the commensal systems of GMRT and uGMRT, but the detection rate is maximum ($\sim 8.26\times10^8$ FRBs per day) for CHIME with IA beam forming mode. However, OWFA and the commensal systems of GMRT, uGMRT and CHIME operate in different frequency ranges and hence complementary.

\begin{table*}[htb]
\tabularfont
\caption{The number of FRBs expected to be detected per day by considering the Dirac delta function as the energy distribution function of FRBs and the mean value of $\alpha$ with error for the range $-5\leq\alpha\leq0$, where $E_\nu\propto\nu^{\alpha}$. Two scattering models and FRB signal without scattering have been considered for the detection rate comparison. Here IA, CA-SB, CA-MB and MIA denote incoherent, coherent single beam, coherent multiple beam and multiple incoherent beam formations respectively. The symbols $\nu$, $B$, $F_{\rm th}$ and FoV denote observational frequency of the telescope, bandwidth of the observation, threshold fluence of FRB required for the detection and the field of view of the telescope respectively.}
\begin{center}
\Rotatebox{90}{%
\begin{tabular}{l|ccccccccc}
\topline
Telescope & System & $\nu$ & $B$ & Beam & $F_{\rm th}$ & FoV & \multicolumn{3}{c}{FRB Detection Rate (${\rm day}^{-1}$)} \\
 &  & (MHz) & (MHz) & Formation & (Jy ms) & (${\rm deg}^2$) & Scattering Model I & Scattering Model II & Without Scattering \\
\midline
    & LS  &  $326.5$   & $4$   & IA & $6.88$  & $0.52$  & $(2.37\pm1.09)\times10^5$  & $(1.17\pm0.64)\times10^6$  & $(2.11\pm0.23)\times10^6$  \\
\cline{2-10}
    &     &     &    & IA & $22.80$  & $24.11$  & $(7.19\pm1.29)\times10^6$  & $(3.34\pm0.91)\times10^7$  & $(8.95\pm0.32)\times10^7$  \\    
    & PI  &  $326.5$   & $19.2$  & CA-SB & $3.61$  & $0.60$  & $(3.43\pm0.98)\times10^5$  & $(1.68\pm0.53)\times10^6$  & $(3.12\pm0.10)\times10^6$ \\
OWFA & &  &   & CA-MB & $1.08$  & $24.11$  & $(1.92\pm0.82)\times10^7$  & $(8.93\pm0.31)\times10^7$  & $(1.38\pm0.13)\times10^8$ \\
\cline{2-10}
    &  &        &  & IA & $38.10$  & $143.34$  & $(3.55\pm1.42)\times10^7$  & $(1.61\pm1.03)\times10^8$  & $(4.33\pm0.48)\times10^8$ \\
    & PII & $326.5$ & $38.4$ & CA-SB & $2.34$  & $0.55$  & $(3.49\pm0.92)\times10^5$  & $(1.69\pm0.45)\times10^6$  & $(2.80\pm0.10)\times10^6$ \\
    & &  &  & CA-MB & $0.70$  & $143.34$  & $(1.29\pm0.77)\times10^8$  & $(5.77\pm0.22)\times10^8$  & $(8.26\pm0.13)\times10^8$ \\
\hline
    & &  &  & IA & $2.25$  & $1.62$  & $(3.60\pm0.95)\times10^5$  & $(1.73\pm0.48)\times10^6$  & $(2.93\pm0.10)\times10^6$ \\
    & Band $S1$ & $300$ & $32$ & CA-SB & $0.41$  & $5.25\times10^{-6}$  & $(5.71\pm0.74)\times10^3$  & $(2.44\pm0.16)\times10^4$  & $(3.32\pm0.10)\times10^4$ \\
GMRT & &  &  & MIA & $1.17$  & $1.62$  & $(4.32\pm0.87)\times10^5$  & $(2.02\pm0.37)\times10^6$  & $(3.12\pm0.12)\times10^6$ \\  
\cline{2-10}
    & &  &  & IA & $2.25$  & $0.72$  & $(4.12\pm0.73)\times10^4$  & $(2.00\pm0.27)\times10^5$ & $(3.41\pm0.20)\times10^5$ \\
    & Band $S2$ & $450$ & $32$ & CA-SB & $0.41$  & $2.33\times10^{-6}$  & $(3.59\pm0.55)\times10^3$ & $(1.57\pm0.10)\times10^4$ & $(2.15\pm0.13)\times10^4$ \\    
    & &  &  & MIA & $1.17$  & $0.72$  & $(5.01\pm0.66)\times10^4$ & $(2.38\pm0.12)\times10^5$ & $(3.67\pm0.18)\times10^5$ \\
\hline
    & &  &  & IA & $5.21$  & $4.26$  & $(5.19\pm1.46)\times10^5$  & $(2.52\pm0.77)\times10^6$ & $(4.27\pm0.41)\times10^6$ \\    
    & Band $2$ & $185$ & $130$ & CA-SB & $0.95$  & $1.38\times10^{-5}$  & $(8.35\pm1.12)\times10^3$ & $(3.67\pm0.52)\times10^4$ & $(5.06\pm0.09)\times10^4$ \\    
    & &  &  & MIA & $2.71$  & $4.26$  & $(6.32\pm1.31)\times10^5$ & $(2.99\pm0.66)\times10^6$ & $(4.64\pm0.29)\times10^6$ \\
\cline{2-10}
    & &  &  & IA & $0.90$  & $1.04$  & $(3.79\pm0.78)\times10^6$ & $(1.73\pm0.22)\times10^7$ & $(2.30\pm0.12)\times10^7$ \\    
    & Band $3$ & $375$ & $250$ & CA-SB & $0.16$  & $3.36\times10^{-6}$  & $(5.83\pm0.59)\times10^3$ & $(2.24\pm0.11)\times10^4$ & $(2.63\pm0.10)\times10^4$ \\    
uGMRT& &  &  & MIA & $0.47$  & $1.04$  & $(4.52\pm0.70)\times10^6$ & $(1.95\pm0.08)\times10^7$ & $(2.46\pm0.14)\times10^7$ \\
\cline{2-10}
    & &  &  & IA & $0.69$  & $0.30$  & $(1.94\pm0.37)\times10^5$ & $(9.01\pm0.31)\times10^5$ & $(1.15\pm0.28)\times10^6$ \\ 
    & Band $4$ & $700$ & $300$ & CA-SB & $0.13$  & $9.64\times10^{-7}$  & $(2.94\pm0.24)\times10^3$ & $(1.17\pm0.26)\times10^4$ & $(1.35\pm0.18)\times10^4$ \\    
    & &  &  & MIA & $0.36$  & $0.30$  & $(2.31\pm0.32)\times10^5$ & $(1.02\pm0.35)\times10^6$ & $(1.25\pm0.23)\times10^6$ \\
\cline{2-10}
    & &  &  & IA & $0.63$  & $0.09$  & $(1.85\pm0.03)\times10^4$ & $(8.82\pm0.60)\times10^4$ & $(1.05\pm0.45)\times10^5$ \\
    & Band $5$ & $1250$ & $400$ & CA-SB & $0.12$  & $3.02\times10^{-7}$  & $(1.55\pm0.13)\times10^3$ & $(6.34\pm0.36)\times10^3$ & $(7.04\pm0.29)\times10^3$ \\    
    & &  &  & MIA & $0.33$  & $0.09$  & $(2.23\pm0.07)\times10^4$ & $(1.01\pm0.49)\times10^5$ & $(1.16\pm0.37)\times10^5$ \\
\hline
    & & & & IA & $5.80$  & $132.00$  & $(1.79\pm0.82)\times10^8$ & $(8.93\pm0.32)\times10^8$ & $(1.17\pm0.11)\times10^9$ \\    
CHIME& $-$ & $600$ & $400$ & CA-SB & $0.16$  & $0.29$  & $(1.10\pm0.37)\times10^6$ & $(4.36\pm0.24)\times10^6$ & $(4.83\pm0.18)\times10^6$ \\

\hline
\end{tabular}
}%
\end{center}
\label{tab:1}
\end{table*}

The quantity $F_{th}$ is an important parameter of FRB detection. We can detect a FRB if and only if the fluence of this event is larger than the value of $F_{th}$ for a given radio telescope. For OWFA, it is found that we can detect bright FRBs by using the IA beam forming mode, whereas CA-MB mode can be used to detect variety of FRBs. Similarly for the two commensal systems of GMRT and the four frequency bands of uGMRT, we can detect variety of FRBs by using MIA beam forming mode, whereas IA mode can be used to detect only bright FRBs. In contrast, CHIME can only detect bright FRBs.

The localization of the event depends on the field-of-view (FoV) of the telescope. A telescope with large field-of-view can localize an event poorly in compare to the same for a telescope with small field-of-view. From Table 1, it is found that the localization of FRBs is much better for the two commensal systems of GMRT and the four frequency bands of uGMRT with CA-SB beam forming mode in compare to others. The localization of the event is much poorer for OWFA, where the dipoles are aligned along the focal line of the cylindrical reflector and hence it can only localize the event along a straight line in north-south direction. However as I have mentioned earlier, the detection probability of FRBs largely depends on the product of field-of-view and sensitivity of the telescope and hence OWFA is capable to detect a large number of FRBs in compare to the same for others mentioned in Table 1. In a brief, we can detect a large number of FRBs with poor localization by using OWFA, whereas the two commensal systems of GMRT and the four frequency bands of uGMRT can be used to detect a comparatively less number of FRBs with better localization. For CHIME, the localization of detected FRBs (field of view = $132\;{\rm deg}^2$) is quite poor in comparison to OWFA, the commensal systems of GMRT and uGMRT. 

\section{Summary and Conclusion}
Fast Radio Bursts (FRBs) are short duration highly energetic dispersed radio pulses. In this review article, I have summarized our predictions of detecting FRBs using OWFA (Bhattacharyya {\em et al.}, 2017) and uGMRT (Bhattacharyya {\em et al.}, 2018) with different beam forming modes. We used the model prescribed by Bera {\em et al.} (2016) for those predictions. We considered the energy spectrum of FRBs as a power law and the energy distribution of FRBs as a Dirac delta function and Schechter luminosity function with both positive and negative exponents. We also considered two scattering models prescribed by Bhat {\em et al.} (2004) and Macquart \& Koay (2013) and FRB pulse without scattering as a special case for the prediction of FRB pulse width.

We have first discussed our predictions of FRB detection rate for the Ooty Wide Field Array (OWFA). OWFA is an upgraded version of the Ooty Radio Telescope (ORT). ORT has a long cylindrical reflector of dimension $530{\rm m}\times 30{\rm m}$,which contains $1056$ half wavelength linear dipoles along the focal line of the reflector. The signals from these dipoles are combined different way and we have discussed our predictions for the old analogue beam forming network, Legacy system, and the upcoming Phase I and II and compared the results with UTMOST and CHIME. In this work, we considered three kind of beam formations; incoherent (IA), coherent single (CA-SB) and coherent multiple (CA-MB) beam formations. We found that we can expect to detect $\sim 10^5-10^8$ FRBs per day by using OWFA with fluence $F\geq0.7\;{\rm Jy\,ms}$. 

We have next discussed our predictions of FRB detection rate for the upgraded Giant Metrewave Radio Telescope (uGMRT) with three kind of beam formations; incoherent (IA), coherent single (CA-SB) and multiple incoherent (MIA) beam formations. The uGMRT is an upgraded version of the GMRT, which has 30 parabolic dishes having $45$ m diameter each and they are distributed in a Y shaped pattern with a shortest baseline of $200$ m and a longest baseline of $25$ km. Each dish has five prime focus feeds having five discrete operational frequencies centered at $150\,{\rm MHz}$, $235\,{\rm MHz}$, $325\,{\rm MHz}$,  $610\,{\rm MHz}$ and $1280\,{\rm MHz}$ with a maximum backend instantaneous frequency bandwidth of $32\,{\rm MHz}$. uGMRT will provide significantly large instantaneous bandwidths with four operational frequencies centered at $185\,{\rm MHz}$, $375\,{\rm MHz}$, $700\,{\rm MHz}$ and $1250\,{\rm MHz}$ with a wide variation of bandwidths from $130\,{\rm MHz}$ to $400\,{\rm MHz}$. In this work we also considered the proposed commensal system for the GMRT for transient search. 
We found that we expect to detect $\sim 10^3-10^6$ FRBs per day with fluence $F\geq0.41\;{\rm Jy\,ms}$ for the commensal system of GMRT and $\sim 10^5-10^7$ FRBs per day with fluence $F\geq0.12\;{\rm Jy\,ms}$ for the four frequency bands of uGMRT respectively. Further it is found that OWFA and the lower frequency bands of GMRT and uGMRT can detect bright FRBs only, whereas the higher frequency bands of GMRT and uGMRT can be used to detect variety of FRBs. It is also found that we can detect a large number of FRBs with poor localization by using OWFA, whereas the two commensal systems of GMRT and the four frequency bands of uGMRT can be used to detect a comparatively less number of FRBs with better localization.

However there are some uncertainties and limitations in our predictions. The scattering mechanism in the intervening medium is still unknown. Further, there is no unique and direct way to estimate the spectral index of FRBs. Moreover, the energy distribution function of FRBs is another important unknown quantity, and we have considered two possible energy distribution models  in this analysis. The detection of a large number of FRBs in future will help us to constrain these uncertainties and refine the FRBs models.

\end{document}